\documentclass[prl, 
    floatfix,
    tightenlines,
    aps,
    showpacs,
    nofootinbib,
    longbibliography,
    twocolumn,
    10pt]{revtex4-2}

\usepackage{amssymb, amsmath, mathtools}
\usepackage{dsfont}

\usepackage[english]{babel}
\usepackage[utf8]{inputenc}
\usepackage{blindtext}
\usepackage{graphicx}

\usepackage{physics}

\usepackage{xcolor}
\usepackage[normalem]{ulem}
\usepackage{float}
\usepackage{pgfplots}
\renewcommand{\vec}[1]{\boldsymbol{#1}}

\newcommand{\ssm}{\rm \scriptscriptstyle}

\renewcommand{\phi}{\varphi}

\usepackage{hyperref}
\hypersetup{colorlinks = true, linkcolor=magenta,citecolor=blue, urlcolor=magenta, bookmarksnumbered =  true}
\pgfplotsset{compat=1.18}
\begin{document}

\title{Partial suppression of magnetism in the square lattice SU(3) Hubbard model}

\author{Samuel Bird}
\email{sabird@phys.ethz.ch}
\author{Sebastian Huber}
\author{Jannes Nys}
\affiliation{
Institute for Theoretical Physics, ETH Zurich, 8093 Z\"urich, Switzerland
}

\begin{abstract}
The ${\rm SU}(N)$ Hubbard model is a natural extension of the ${\rm SU}(2)$ model. However, even the $N=3$ case remains poorly understood. We report a substantially new ground-state phase diagram of the square lattice ${\rm SU}(3)$ Fermi-Hubbard model. Using a backflow ansatz, we identify strong signatures of a Mott transition and a subsequent magnetic transition, and the suppression of a previously predicted magnetic phase. We study the hole-doped model, identifying a transition from magnetic to paramagnetic behavior in the strong-coupling regime. Our findings offer a qualitatively new ground state picture. More broadly, our work suggests a path to study general ${\rm SU}(N)$ Hubbard models with arbitrary filling and geometry.
\end{abstract}

\maketitle

\textbf{\textit{Background. }}The single-band ${\rm SU}(2)$ symmetric Fermi-Hubbard model (FHM) \cite{HubbardProc1963, GutzwillerPR1965} is the prototypical lattice model for strongly correlated electronic materials. It describes a Mott insulator and itinerant antiferromagnetism (AF), reducing to the Heisenberg model in the strong coupling limit, and, upon doping, it can display $d$-wave superconductivity, a pseudo-gap regime, strange metallicity, and more \cite{HirschPRB1985, Bednorz1986, WhitePRB1989, rasettiWorldSci1991, scalapinoProcISP1994, scalapinoJPCS1995, fazekas1999lecture, LeBlanc2015, Zheng2017,Qin2020, Jiang2022, Arovas2022}. 

The $\left(N>2\right)$ ${\rm SU}(N)$ symmetric FHM is a natural extension of the ${\rm SU}(2)$ case. The higher spin symmetry allows one to disentangle the effects of the strong interaction from other effects particular to the ${\rm SU}(2)$ model -- such as the perfect nesting of the Fermi surface -- and can introduce frustration. Notable direct applications are systems such as graphene with ${\rm SU}(4)$ spin-valley symmetry, multi-orbital models used to describe transition metal oxides \cite{LiPRL1998,TokuraScience2000,HafezTorbati2018}, or 2D Moiré materials \cite{WuPRL2018, PanPRR2020, KuhlenkampPRX2024}. Moreover, the experimental realization of ${\rm SU}(N)$ Fermi-Hubbard models is now possible using ultra-cold alkaline earth atoms (AEA) in optical lattices. They possess a tunable nuclear spin degree of freedom, which makes it possible to realize ${\rm SU}(N)$ FH models with $N$ ranging from two to ten \cite{HonerkampPRL2004, Hermele2009,Gorelik2009,BlochNatPhys2012, SotnikovPRA2014, Bohrdt2021,Gorshkov2010,Hermele2011,TaieNatPhys2012,Cazalilla_2014,OzawaPRL2018,HazzardPRA2021,TaieNatPhys2022,HazzardPRL2024, IGPadillaIOP2024}.

\begin{figure}[t]
    \centering
    \includegraphics{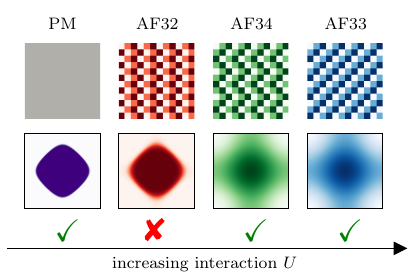}
    \caption{\textbf{Phase diagram.} An illustration of the different magnetic phases obtained in this work. The top row shows an illustration of the classical flavor states, the bottom panels sketch the momentum distribution in each phase. Only the paramagnet (PM) is characterized by a sharp Fermi surface.}
    \label{fig:phases}
\end{figure}

Despite previous theoretical and experimental progress, much remains to be understood about the ground state and low temperature phase diagrams of the ${\rm SU}(N)$ FH models \cite{CapponiAOP2016,ManmanaPRA2011,XuPRL2018, AssarafPRB1999, AssaadPRB2005, CaiPRB2013, BlumerPRB2013, WangPRL2014}. Unlike the square ${\rm SU}(2N)$ FHM at $1/2$ filling, the ${\rm SU}(3)$ model possesses a fermionic sign problem that makes low-temperature intermediate to strong correlation regimes hard to access with conventional quantum Monte Carlo (QMC) methods \cite{LohPRB1990, mondainiScience2022, troyerPRL2005, umrigarPRL2007}. In the very strong correlation limit, the Hubbard model maps onto a ${\rm SU}(3)$ Heisenberg model. Numerical studies of this model suggest a ground state possessing 3-sublattice (3SL) ``stripe" antiferromagnetic order, emerging also via ``order-by-disorder" in flavor-wave theory \cite{TothPRL2010, BauerPRB2012, RomenPRR2020}. Recent constrained-path auxiliary-field QMC (AFQMC) \cite{HazzardPRR2023} and determinant QMC \cite{HazzardPRA2023} studies on the ${\rm SU}(3)$ square FHM suggest that two more AF states lie between this strong-coupling state and a weak-coupling paramagnetic metal. Finally, both the effects of spin-imbalance \cite{Tusi_2022,zhangArxiv2025}, single-hole doping in the $t-J$ limit \cite{SchlomerPRB2024}, and other geometries \cite{Nie2017,Chung2019,PerezRomero2021,BohlerArxiv2025} have also been investigated.

In this work, we use variational Monte Carlo (VMC) to study the ground state phase diagram of the ${\rm SU}(3)$ square FHM. A key utility in our work is the freedom to modify the nodal surface of the wave function during optimization. We can target a full range of interaction strengths and doping concentrations, all within the same scheme and for large systems. The scheme is also readily applicable to higher ${\rm SU}(N)$.

At unit density, we find a qualitatively different ground state phase diagram than predicted in previous studies \cite{HazzardPRR2023, HazzardPRA2023, zhangArxiv2025}. We observe a total suppression of one of the predicted intermediate AF phases, illustrated in Figure \ref{fig:phases}. The nature of the magnetic transition, originally believed to be second-order, with magnetic order continuously growing from zero, changes to a first-order transition from a paramagnetic state to a state with finite magnetic order. We attribute this difference to the presence of a fixed nodal surface in previous studies. We find strong signatures of the Mott transition, through the number of doubly occupied sites $D$ and the momentum-space density $n(\vec{k})$, which present discontinuous behavior between the metallic and insulating regimes but not between different magnetic insulating states. We find a critical interaction strength of $U\approx 7.375\,t$, much higher than previously believed. We also study the effects of doping the ${\rm SU}(3)$ FHM with holes. We conducted a sweep of doping concentrations and identified the crossover from magnetism to paramagnetism, comparing our results at different couplings.

\textbf{\textit{Model and previous results. }}The ${\rm SU}(N)$ FHM, with nearest-neighbor hopping, is described by the tight-binding Hamiltonian
\begin{equation}
\begin{aligned}
\hat{H} = -t\sum_{\langle ij\rangle}\sum_{\sigma=1}^N\hat{c}^\dagger_{i\sigma}\hat{c}_{j\sigma} + \text{h.c.} + U\sum_{i}^{N_s}\sum_{\sigma <\tau}^N\hat{n}_{i\sigma}\hat{n}_{i\tau},
\end{aligned}
\end{equation}
with a Hubbard interaction term proportional to $U$. Greek letters denote the ${\rm SU}(N)$ symmetric flavor, $t$ is the hopping amplitude between nearest neighbors $\langle i,j\rangle$. We focus on the case of $N=3$ flavors and a square lattice. We take a fixed number of fermions and assume flavor balance for simplicity. All simulations are performed on a $12\times 12$ super-cell with $N_s=144$ sites.

The latest ground state studies \cite{HazzardPRR2023, zhangArxiv2025} of the ${\rm SU}(3)$ FHM at unit charge density predict a Mott transition at finite $U$ from a paramagnetic metal (PM) to an antiferromagnetic state that breaks translation symmetry with the ordering vector $(3,2)$, corresponding to periodicity 3 in one direction and periodicity 2 in the other, which we denote AF32. As $U$ increases, this is followed by a transition to an intermediate AF state with periodicity $(3,4)$ AF34 and finally to the AF state with periodicity $(3,3)$ AF33, corresponding to the diagonal color stripes or 3-sublattice order of the $SU(3)$ Heisenberg model \cite{TothPRL2010, BauerPRB2012}. These phases are depicted in Fig.~\ref{fig:phases}.

\textbf{\textit{Methodology. }} We perform VMC simulations to approximate the ground state wave function by a variational trial wave function $\ket{{\psi_{\vec{\theta}}}}$ \cite{Becca_Sorella_2017}.  

In this work, we use two different variational ansätze: Slater-Jastrow with and without backflow. The base ansatz is a Slater determinant with a two-body density-density Jastrow correlator. The Slater determinant (SD) allows for ${\rm SU}(3)$ symmetry breaking, analogous to unrestricted Hartree-Fock \cite{XuIOP2011}. The Slater determinant wave function is given by the determinant of a Slater matrix $M$ built from single-particle orbitals evaluated at fermionic coordinates 
\begin{equation}
    \begin{aligned}
        M&=\begin{pmatrix}
\phi_1(\boldsymbol{r}_1, \sigma_1)&&...&&\phi_1(\boldsymbol{r}_{N_f},\sigma_{N_f}) \\
\vdots&&\ddots&&\vdots \\
\phi_{N_f}(\boldsymbol{r}_1, \sigma_1)&&\dots&&\phi_{N_f}(\boldsymbol{r}_{N_f},\sigma_{N_f}) \\
\end{pmatrix}.
    \end{aligned}
\end{equation}
We choose $M$ to be block diagonal in the flavor degree of freedom. 
\begin{equation}
\begin{aligned}
\Psi_{\ssm SD}&=\det M =\prod_\sigma \det M_\sigma \\
\end{aligned}
\end{equation}

\begin{figure*}[t]
    \centering
    \includegraphics{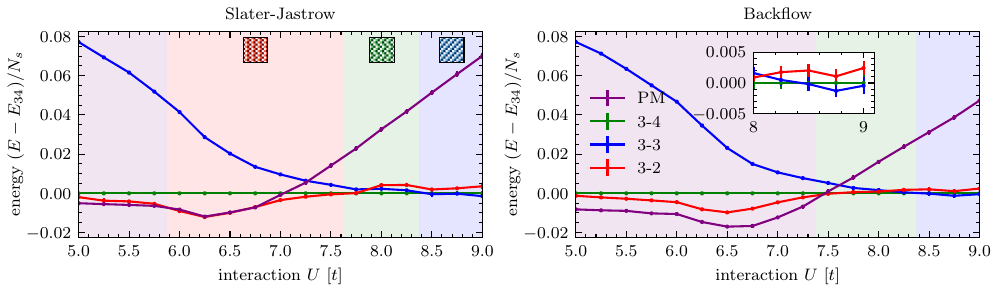}
    \caption{\textbf{Energy comparison.} The energy per site for each of the PM, AF32, AF34, and AF33 states, calculated at each value of $U$. Energy is given relative to the energy of the AF34 state. Error-bars are smaller than the markers where not visible. Left: Results obtained with the Slater-Jastrow ansatz. Insets show the different magnetic order patterns of each phase.  Right: Results obtained with the backflow ansatz. Inset zooms into the strong-coupling region.}
    \label{fig:energy}
\end{figure*}

The translation-invariant two-body Jastrow correlator is then applied. In the occupation basis,
\begin{equation}
    \begin{aligned}
        \Psi_{\ssm SJ}\left(\{n_{i\sigma}\}\right) & = \mathcal{J}\left(\{n_{i\sigma}\}\right)\Psi_{\ssm SD}\left(\{n_{i\sigma}\}\right) \\
        \mathcal{J}\left(\{n_{i\sigma}\}\right) & = \exp{\sum_{i,j=1}^{N_s}\sum_{\tau\geq\sigma}^3J_{\sigma,\tau}\left(|\vec{r}_i-\vec{r}_j|\right)n_{i\sigma}n_{j\tau}}, 
    \end{aligned}
\end{equation}
where $J_{\sigma,\sigma}$ is the same for all $\sigma$ and $J_{\sigma,\tau}$ is the same for all $\sigma\neq\tau$, preserving ${\rm SU}(3)$ symmetry. The matrix $M_{\mu\nu}$ and the Jastrow factors $J_{\sigma,\tau}\left(r\right)$ are the variational parameters in our model.

Further correlations are added to the Slater-Jastrow ansatz by backflow \cite{TocchioPRB2008,TocchioPRB2011}. Backflow introduces a many-body fermion configuration dependence into the Slater determinant and variationally modifies the wave function's nodal surface. We use a nearest-neighbor orbital backflow \cite{TocchioPRB2011}, which we generalize to the ${\rm SU}(N)$ case, with two variational parameters $\eta, \epsilon$, 
\begin{equation}
    \begin{aligned}
    \tilde{\phi}_\mu (\vec{r}_{i}, \sigma_i) &=  \phi_{\mu}(\vec{r}_{i}, \sigma_i) \\
    &+ \epsilon \hat{D}_i \min\bigg(\sum_{j \in nn} \hat{H}_j,1\bigg)\phi_\mu(\vec{r}_i, \sigma_i) \\
    &+ \eta \sum_{j \in nn} \hat{D}_i \hat{H}_j \phi_\mu(\vec{r}_{j},  \sigma_i). 
    \end{aligned}
\end{equation}
The orbitals $\phi_\mu\rightarrow\tilde{\phi}_\mu$ are augmented by the presence of doubly occupied and empty sites through the projection operators $\hat{D}_i = \hat{P}_{i,2},\,\hat{H}_i = \hat{P}_{i,0}$, where $\hat{P}_{i,n}$ is the projector onto the manifold of states with $n$ fermions on site $i$.

\begin{figure}[b]
    \centering
    \includegraphics{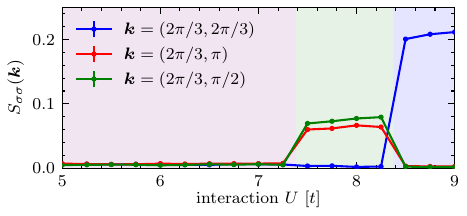}
    \caption{\textbf{Structure factor.} Structure factor $S(\vec{k})$ evaluated at the 3 representative momenta, in the lowest energy backflow state at each $U$ (indicated by the background shade). Error-bars are smaller than the markers where not visible.}
    \label{fig:spin}
\end{figure}

We first optimize our unrestricted Slater determinants at each $U$, ensuring that they are converged, before applying Jastrow and then backflow corrections in turn. A generic difficulty is that there are many states very close in energy as a result of frustration \cite{TothPRL2010}. To aid convergence to symmetry-broken states, we thus use a pinning-field approach, inserting external potentials over the whole super-cell to explicitly break translation and $SU(3)$ symmetry. We then turn them off during optimization of the Slater determinant. In this way, the state is initialized in a symmetry-broken subspace, but as the pinning field is turned off, the state is free to explore the full variational Hilbert space. We do this separately for each of the predicted magnetic orders, and optimize without a pinning field to obtain the paramagnetic state. We then verify that the final state has the corresponding magnetic correlations. 

Minimization of the variational energy is performed by Stochastic Gradient Descent (SGD) and minimal-step Stochastic Reconfiguration (minSR) \cite{ChenNature2024,RendeCommPhys2024}.

\textbf{\textit{Results and discussion.} } The left panel of Fig.~\ref{fig:energy} shows the phase diagram obtained using the Slater-Jastrow ansatz at unit density. It accurately reproduces the AFQMC phase diagram \cite{HazzardPRR2023}, capturing the low-coupling paramagnetic phase, the strong-coupling AF33 state, and the intermediate AF32 and AF34 states, with similar critical values of $U$.

The picture changes when we allow the nodal surface to change, using the backflow ansatz as shown in the right panel of Fig.~\ref{fig:energy}. We find that backflow corrections universally favor the PM state over AF32 in the lower-coupling regime. In the Slater-Jastrow phase diagram, one can see that the PM and AF32 states are very close in energy, so this highlights the role of the backflow corrections in distinguishing the states. The onset of magnetization is much higher in $U$ than predicted by Slater-Jastrow, but the onset of AF34 occurs at a lower $U_{c1}\approx 7.375\,t$. The transition to AF33 seems to be unchanged, around $U_{c2}\approx 8.375\,t$. In the strong-coupling regime that we have studied, the AF33 and AF34 states are actually within error of each other. Further discussion of the AF32 phase can be found in the Supplementary Material \cite{SM}.

Antiferromagnetic correlations are analyzed by calculating the ${\rm SU
}(3)$-symmetric flavor correlation function $S(\vec{r}_i-\vec{r}_j) = \langle\sum_\sigma \hat{n}_{i\sigma}\hat{n}_{j\sigma}\rangle$ and the structure factor $S(\vec{k}) = (1/N_s)\sum_{\vec{r}}e^{i\vec{k}\cdot\vec{r}}S(\vec{r})$, cf. Fig.~\ref{fig:spin}. At the transition to the AF34 phase, finite peaks appear in the structure factor at $\vec{k}=(2\pi/3,\pi)$ and $\vec{k}=(2\pi/3,\pi/2)$. At the transition to AF33, both peaks vanish discontinuously and a new peak emerges at $\vec{k}=(2\pi/3,2\pi/3)$. Every transition appears to be first-order.

The identification of a Mott transition is subtle. We examine two signatures: the number of doubly occupied sites $D=(1/N_s)\langle\sum_{i,\sigma\neq\tau}\hat{n}_{i\sigma}\hat{n}_{i\tau}\rangle$, relevant for frustrated models \cite{TocchioPRB2011}, and the momentum-space density $n(\vec{k})=\langle 1/3\sum_\sigma \hat{n}_\sigma(\vec{k})\rangle$, whose discontinuity determines the quasiparticle weight $Z$ in a metallic state. Note that $\hat{n}_\sigma(\vec{k}) =1/N_s\sum_{ij}e^{i(\vec{r}_i-\vec{r}_j)\cdot\vec{k}}\hat{c}^\dagger_{i\sigma}\hat{c}_{j\sigma}$.

\begin{figure}[t]
    \includegraphics{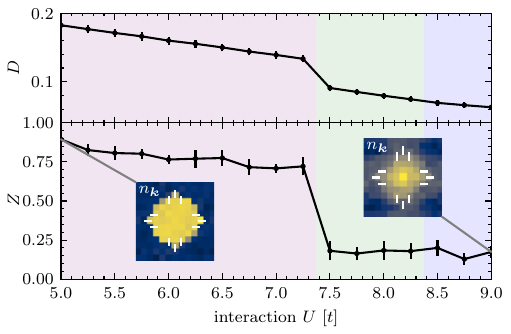}
    \caption{\textbf{Number of doubly occupied sites and quasiparticle weight.} Top: The density of doubly occupied sites $D$, evaluated in the lowest energy state at each $U$. Error-bars are smaller than the markers where not visible.  Bottom: The quasiparticle weight $Z$ at each $U$. Insets show the Fermi sea at two representative points, with white lines indicating the positions and directions around which we calculate the discontinuity corresponding to the quasiparticle weight.}
    \label{fig:nk}
\end{figure}

For the backflow wave function, $D$ decreases smoothly in the PM phase and then discontinuously jumps down at the transition to AF34 as shown in Fig.~\ref{fig:nk}. This is a signature of a Mott transition at the critical point $U_{c1}\approx7.375\,t$. Note that $D$ continues to decrease smoothly across the transition to AF33. This discontinuity has not been observed in previous $T=0$ studies to the best of our knowledge.

The profile of $n(\vec{k})$ as a function of $U$ shows the presence of a discontinuous jump around the Fermi surface in a metallic phase and a smoothing of the Fermi surface in an insulating phase, cf. Fig.~\ref{fig:nk}. We estimate the size of the jump $Z$ by calculating the finite difference across the indicated white bars. While this quantity suffers significantly from finite size effects, its behavior as a function of $U$ corroborates the identification of the Mott transition using the double occupancy $D$ at $U_{c1}\approx 7.375\,t$, previously anticipated to be around $5.5\,t$.

\begin{figure}[tb]
    \centering
    \includegraphics{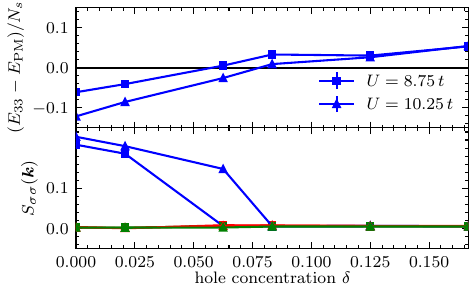}
    \caption{\textbf{Doped energy diagram and structure factor.} Top: Energy per site calculated of the AF33 state relative to the PM state, at a range of hole concentrations, and for both $U=8.75\,t$ and $U=10.25\,t$. Calculations were performed using the Slater-Jastrow ansatz. Bottom: Corresponding flavor structure factor, with the same color code as Figure \ref{fig:spin}. Error-bars are smaller than the markers where not visible.}
    \label{fig:dope}
\end{figure}

We examined the effect of hole doping in the strong-coupling AF33 phase. Figure~\ref{fig:dope} shows the energy of the AF33 state relative to the paramagnetic state, as a function of the hole concentration $\delta=(N_s-N_f)/N_s$. A transition into a regime where the PM state has a lower energy than AF33 occurs around the hole concentration $\delta =0.0625$, for $U=8.75\,t$. This loss of preferability of the AF33 state at low doping is not surprising given the proximity of $U$ to the AF33 transition. At higher coupling $U=10.25\,t$, we can see that the transition occurs at a higher hole concentration, suggesting a more robust magnetism at stronger couplings. 

The bottom panel of Figure \ref{fig:dope} correspondingly shows the decay of the AF33 component of the flavor susceptibility structure factor as the hole concentration increases, dropping to zero at the transition to paramagnetism. 

\textbf{\textit{Conclusion. }}Using a backflow ansatz, we have established a new phase diagram of the ${\rm SU}(3)$ Fermi-Hubbard model. We have analyzed the magnetic properties of the phase diagram by studying the flavor correlation structure factor and identified the Mott transition using the number of doubly occupied sites and the quasiparticle weight derived from the momentum-space density. Our results differ from previous studies, which we were able to reproduce using a simpler ansatz, in the total suppression of the previously proposed AF32 magnetic phase and the expansion of the AF34 phase, which we attribute to the modification of the nodal surface of the wave function. Our results also offer the first strong nonmagnetic signatures of the proposed Mott transition. Finally, we studied the doped model and identified the crossover between magnetism and paramagnetism as holes are added, verifying that the AF33 magnetic order becomes more robust as $U$ increases. The application to the doped system highlights the flexibility of this method.

The demonstrated flexibility of this method allows for the investigation of other phenomena: exotic spin-charge orders and superconductivity upon doping of the ${\rm SU}(3)$ model, dimerized and plaquette magnetism in the ${\rm SU}(4)$ model and higher $N$ \cite{CorbozPRL2011, Nataf2016, Unukovych}, spin liquids in Fermi-Hubbard models \cite{Hermele2011}, and bosonization in the ${\rm SU}(N\rightarrow\infty)$ limit \cite{SongPRX2020}. Finally, the method is readily extendable to more complicated ansätze. Our orbital backflow corrections only introduce an additional $2$ parameters on top of the $>20'000$ parameters of the Slater-Jastrow, and already we see significant changes in the phase diagram. A natural next step is a complete phase diagram calculation using neural quantum states \cite{Carleo_2017, MedvidovicEPJ2024}, which could allow for an even more flexible modification of the nodal surface, which we would expect to strengthen our conclusions and give quantitative improvements to the estimation of critical points.

While finalizing our manuscript, we became aware of a related study using an iPEPS variational wave-function \cite{kleijweg2025zigzagantiferromagnetssu3hubbard}.

\textbf{Acknowledgments. }Our simulations were carried out with the help of NetKet \cite{NetKet}.
\bibliographystyle{phd-url}
\bibliography{ref}

\end{document}


\title{Supplementary Material for: Partial suppression of magnetism in the square lattice SU(3) Hubbard model}
\author{Samuel Bird}
\author{Sebastian Huber}
\author{Jannes Nys}
\affiliation{
Institute for Theoretical Physics, ETH Zurich, 8093 Z\"urich, Switzerland
}

\date{June 2025}

\maketitle

\noindent\footnote{Corresponding author: \texttt{sabird@phys.ethz.ch}}

\section{Specific details of the vmc}
We clarify some specific choices made in the ansätze. We chose a form of Slater determinant assuming the Slater matrix to be block diagonal in the flavor degree of freedom, with nonidentical blocks. This allows for spontaneous breaking of the $\rm{SU}(3)$ symmetry. 
\begin{equation}
    \begin{aligned}
        M&=\begin{pmatrix}
\phi_1(\boldsymbol{r}_1, \sigma_1)&&...&&\phi_1(\boldsymbol{r}_{N_f},\sigma_{N_f}) \\
\vdots&&\ddots&&\vdots \\
\phi_{N_f}(\boldsymbol{r}_1, \sigma_1)&&\dots&&\phi_{N_f}(\boldsymbol{r}_{N_f},\sigma_{N_f}) \\
\end{pmatrix} \\
\\
\Psi_{\ssm SD}&=\det M =\prod_\sigma \det M_\sigma \\
    \end{aligned}
\end{equation}
One could further allow for the breaking of flavor sector number conservation using a fully general Slater matrix, but this increases the computational cost of taking the determinant and we found no benefit for this problem. It would also increase the number of parameters from $3\times (48\times 144)$ to $144\times(3\times 144)$. 

Our Jastrow correlator is chosen to introduce charge and spin correlation holes into the wave function. We impose $\rm{SU}(3)$ symmetry by making intra-flavor and inter-flavor Jastrow factors spin independent. 
\begin{equation}
    \begin{aligned}
        \Psi_{\ssm SJ}\left(\{n_{i\sigma}\}\right) & = \mathcal{J}\left(\{n_{i\sigma}\}\right)\Psi_{\ssm SD}\left(\{n_{i\sigma}\}\right) \\
        \mathcal{J}\left(\{n_{i\sigma}\}\right) & = \exp{\sum_{i,j=1}^{N_s}\sum_{\tau\geq\sigma}^3J_{\sigma,\tau}\left(|\vec{r}_i-\vec{r}_j|\right)n_{i\sigma}n_{j\tau}}, 
    \end{aligned}
\end{equation}
It's important to note that the distance here, $|\vec{r}_i-\vec{r}_j|$, is calculated in the Manhattan metric for a square lattice. With periodic boundary conditions, this means that the maximal distance between any 2 points on a $L\times L$ super-cell is $L$. This means that the Jastrow correlator introduces an additional $2\times 12$ variational parameters in the case of a $L=12$.

The total number of variational parameters used in the Slater-Jastrow ansatz is 20'760, at least in the undoped case. For a general $\rm{SU(N)}$, with $N_f$ fermions, flavor balance, and $L=\sqrt{N_s}$, the number of variational parameters is $N\times(N_f/N\times L^2)\,+2\times L = N_f\times N_s+2\times L$, and is independent of the number of flavors $N$. The backflow corrections \cite{TocchioPRB2008,TocchioPRB2011} introduce only two additional variational parameters, increasing the total number to 20'762, or $N_f\times N_s+2\times L + 2$ in the general case.

We optimized our Slater determinant state using Stochastic Gradient Descent for 1500 update steps, checking Energy convergence at the end. It is during this first initialization that we perform our pinning procedure as outlined in the main paper. After this, we turned on the Jastrow correlator and optimized this using minSR until convergence \cite{ChenNature2024,RendeCommPhys2024}, which we found to perform very well and converge more stably than bare Stochastic Reconfiguration. Finally, we turned on the backflow corrections, once again optimizing using minSR until convergence. All final simulations were performed using a learning rate between $10^{-2}$ and $10^{-3}$, a sample size of $2^{12}$, and with a diagonal offset of $10^{-4}$ in the minSR to aid stability as is common practice.  

\section{Discussion of the AF32 phase and Slater-Jastrow results}

We can gain some insight into the AF32 phase that appears at the level of Slater-Jastrow, and has been present in phase diagrams derived from other methods. In particular, we can understand why it tends to appear in the absence of backflow corrections. 

\begin{figure}[H]
    \centering
    \includegraphics[width=0.49\textwidth]{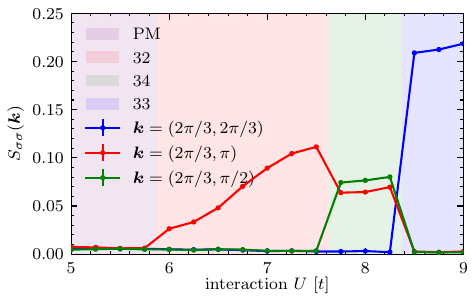}
    \caption{\textbf{Structure factor.}  Structure factor $S(\vec{k})$ evaluated at the 3 representative momenta, in the lowest energy Slater-Jastrow state at each $U$ (indicated by the background shade). Error-bars are smaller than the markers where not visible.}
    \label{fig:spin_sm}
\end{figure}

\begin{figure}[h]
    \centering
    \includegraphics[width=0.49\textwidth]{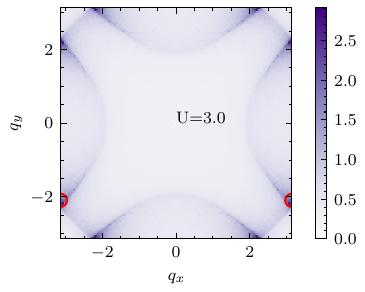}
    \caption{\textbf{RPA flavor susceptibility.} Flavor susceptibility calculated in the 1-loop RPA approximation at low $U=3.0$. Red circles point to $\vec{q}=(2\pi/3,\pi)$, indicating $3-2$ order.}
    \label{fig:rpa_sm}
\end{figure}

Consider the antiferromagnetic flavor structure factor. The structure factor at vector $\vec{k}=(2\pi/3,\pi)$ grows linearly in $U/t$ in the AF32 phase. At the transition to the AF34 phase, the structure factor at $\vec{k}=(2\pi/3,\pi)$ jumps discontinuously and a finite peak at $\vec{k}=(2\pi/3,\pi/2)$ appears. At the transition to AF33, both peaks vanish discontinuously and a new peak emerges at $\vec{k}=(2\pi/3,2\pi/3)$.

The backflow structure factor is similar to the Slater-Jastrow case in the corresponding phases, with discontinuities at every transition (Fig. \ref{fig:spin_sm}), while the linear growth of $S(\vec{k}=(2\pi/3,\pi))$ in AF32 is absent. Thus, every transition appears to be first order, while the transition to AF32 in the Slater-Jastrow case appears to be a continuous transition.

Moreover, one can look at the number of doubly occupied sites. In the Slater-Jastrow result, $N_d$ changes smoothly across the transition to AF32 and then jumps at the transition to AF34. This would seem to suggest that a Mott transition is taking place at the onset of AF34 and not AF32, in agreement with backflow, with the Slater-Jastrow AF32 state being metallic. This is consistent with the behavior of the AF32 structure factor, and the fact that the AF32 and PM states are very close in energy. Note that the backflow appears to further smoothen the jump in $N_d$ across the transition to AF33 compared to pure Slater-Jastrow.

Although our backflow results suggest the predicted AF32 phase disappears from the ground-state phase diagram, it is not surprising that other methods might overestimate magnetic effects. In addition to the Slater-Jastrow picture, we also performed a one-loop random phase approximation (RPA) calculation of the flavor channel susceptibility, which suggests a tendency to $3-2$ magnetic order in the weak-coupling metallic regime (Fig. \ref{fig:rpa_sm}).

The flavor susceptibility $\chi_\mathrm{flavor}$ in the RPA approximation is given by calculating first the Lindhard function $\chi_0$.
\begin{equation}
    \begin{aligned}
        \chi_\mathrm{flavor}\left(\boldsymbol{q}\right) &= \frac{\chi_0\left(\boldsymbol{q}\right)}{1-U\chi_0\left(\boldsymbol{q}\right)}, \\
        \chi_0\left(\boldsymbol{q}\right) &= \frac{1}{N_s}\sum_{\boldsymbol{k}}\frac{n^{(0)}_{\boldsymbol{k}}-n^{(0)}_{\boldsymbol{k}+\boldsymbol{q}}}{\left(\epsilon_{\boldsymbol{k}+\boldsymbol{q}}-\epsilon_{\boldsymbol{k}}\right)},
    \end{aligned}
\end{equation}
in terms of the non-interacting $T=0$ fermionic density function $n^{(0)}_{\boldsymbol{k}} = \Theta(\mu-\epsilon_{\boldsymbol{k}})$, where $\Theta$ is the Heaviside function, and the non-interacting single particle energies $\epsilon_{\boldsymbol{k}}$. This is calculated numerically by discretizing the Brillouin zone. We used a discretization of $121\times121$.

\section{Charge degrees of freedom in the doped model}

\begin{figure}[]
    \centering
    \includegraphics[width=0.49\textwidth]{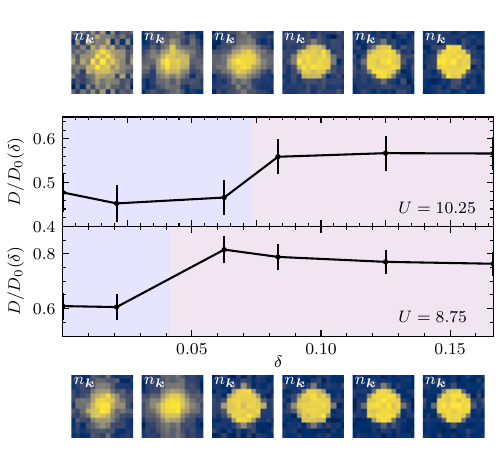}
    \caption{\textbf{Number of doubly occupied sites and $n_k$.} The number of doubly occupied sites and $n_k$ at both $U=8.75\,t$ (top) and $U= 10.25\,t$. The number of doubly occupied sites $D$ is defined relative to $D_0(\delta)$ which is the value of the non-interacting $SU(3)$ Fermi gas on the square lattice at hole concentration $\delta$. }
    \label{fig:doped_charge}
\end{figure}

We have also taken the time to look at the full set of observables in the hole-doped problem. Fig. \ref{fig:doped_charge} shows the density of doubly occupied sites $D/D_0(\delta)$, defined relative to the non-interacting $SU(3)$ Fermi gas value $D_0(\delta)$ to disentangle effects from the interaction and from the pure change in particle number, and the evolution of $n_k$ as the hole concentration increases. Both of these observables describe charge degrees of freedom as opposed to flavor degrees of freedom. 

The non-interacting $\rm{SU}(3)$ Fermi gas value is given by simple combinatorics
\begin{equation}
    \begin{aligned}
    D_0(\delta) = \left(\frac{1}{3}-\frac{\delta}{3}\right)^2,
    \end{aligned}
\end{equation}
in the flavor balanced case with total hole concentration $\delta = (N_s-N_f)/N_s$.

Relative to the Fermi gas value, Fig. \ref{fig:doped_charge} shows an increase in the number of doubly occupied sites as the hole concentration increases, highlighting the difference between the metal and insulator behavior. A similar qualitative shift in sharpness of the Fermi sea in $n_k$ can also be seen.

\bibliography{ref}